# Graphing Website Relationships for Risk Prediction

Identifying Derived Threats to Users Based on Known Indicators

Philip H. Kulp, Capitol Technology University, phkulp@captechu.edu
11301 Springfield Rd, Laurel, MD 20708

Nikki E. Robinson, Capitol Technology University, nerobinson@captechu.edu
11301 Springfield Rd, Laurel, MD 20708

## ABSTRACT

The hypothesis for the study was that the relationship based on referrer links and the number of hops to a malicious site could indicate the risk to another website. We chose Receiver Operating Characteristics (ROC) analysis as the method of comparing true positive and false positive rates for captured web traffic to test the predictive capabilities of our model. Known threat indicators were used as designators, and the Neo4j graph database was leveraged to map the relationships between other websites based on referring links.

Using the referring traffic, we mapped user visits across websites with a known relationship to track the rate at which users progressed from a non-malicious website to a known threat. The results were grouped by the hop distance from the known threat to calculate the predictive rate. The results of the model produced true positive rates between 58.59% and 63.45% and false positive rates between 7.42% to 37.50%, respectively. The true and false positive rates suggest an improved performance based on the closer proximity from the known threat, while an increased referring distance from the threat resulted in higher rates of false positives.

## KEYWORDS

cyber security, graphing database, Receiver Operating Characteristics, Neo4j, website, threat model

# 1  Introduction

The purpose of the research was to test a method of identifying risks to websites based on the relationship to known threats. Since a lag exists between identifying an instance of malware and propagation of the threat notifications, blacklists cannot provide a complete solution to identifying risks [1]. The rationale for the study was to test whether a risk could be predicted for non-malicious websites based on the proximity to malicious sites as determined by known threat indicators. In the current research, we sought a method of enhancing the existing threat indicators with additional metadata about the evaluation of risk based on relationships.

Receiver Operating Characteristics (ROC) analysis [2] was used to evaluate the performance of predicting the risk of visiting websites based on the proximity to known malicious sites. We defined proximity as the number of hops to a known threat. ROC analysis allows researchers to evaluate models in medical trials, intrusion detection systems, and machine learning. Since ROC analysis has been used to evaluate the performance of Intrusion Detection Systems (IDS) and other cybersecurity-related research [3], we concluded the method could provide an acceptable approach for testing binary classification of threats.

ROC analysis can be used to evaluate multiple classes of a single model, and in our study, the classes were defined as the distance (in hops) from the known threat. By delineating threats based on the relationships to other websites, organizations could use the information to determine the threshold for acceptable risk. The acceptable risk would be based on the probability that a user visiting a website may follow a path to malicious content.

# 2  Related Work

Gyongyi, Garica-Molian, and Pedersen performed research [4] on a similar topic as the current study using the equivalent of Google PageRank to determine if a website should be classified as spam. The researchers focused the discussion on a method called TrustRank, which evaluated multiple indicators to assess the trust of a website. Wen, Zhao, and Yan presented similar research [5] as Gyongin et al. except the researchers analyzed the content of pages to determine whether the website should be classified as malicious. While both research studies presented similar topics as ours, the goal of the current study was to track activity across related websites to test the likelihood of predicting a user visiting a malicious website after entering other websites in the relationship.

Rawal, Liang, Loukili, and Duan [6] focused on the research in the prior ten years of cyberattacks, the attacker's patterns, and thus offered a tool to predict the subsequent moves of the attackers. This research aligns with our research in attempting to understand the next move of the attacker, which would provide an advantage to the organization under threat of attack. Rawal et al. proposed a mathematical model to predict cyberattacks. The information used for this model included an analysis of hundreds of cyberattacks from 2000 to 2015. The researchers first identified who the possible attackers would be, including internal and external intruders, then reviewed the research to identify the top attack methods used. The top three attack vectors included Denial of Service (DoS), Cross-site Scripting (XSS), and Brute Force [6]. The results of their research suggest the need for a tool to predict when a user or organization is at risk of a cyberattack.

Chiba, Tobe, Mori, and Goto [1] concurred with Rawal et al. [6] findings that web-based malware attacks are on the rise and considered a severe threat. Chiba et al. tested a method of blacklisting websites using IP characteristics [1]. This study is complementary to our research as we are not attempting to replace blacklists or common cybersecurity tools but to create a model that may aid in identifying malicious websites. Chiba et al. used two Internet Protocol (IP) characteristics of stability and address space characteristics to assist in the blacklisting of websites. The researchers were able to demonstrate the characteristics within the experimental use of their feature extraction model. A key takeaway from this research is the usage of a ROC curve and agreement that there is a need to develop a tool to detect malicious websites. While Chiba et al.

focused on specifically blacklisting of websites, our research focused on determining relationships between malicious websites.

Bruijn and Janssen studied the difficulties in defining a framework for improving the communication of cybersecurity-related issues [7]. The researchers built an evidence-based framework to improve awareness of cybersecurity and improve perspectives for solving problems. Bruijn and Janssen identified four main issues in creating cybersecurity policies; intangible nature, socio-technical dependence, the ambiguous impact of a cyberattack, and defense against cyberattackers. The researcher focused on the inability to communicate risk to users about accessing websites. They also determined the need for creating policies to educate users on website safety. Our research goal was to develop a model which could provide users with risk-based alerts during browsing near malicious websites. The implementation of the model could aid organizations with the creation of policies and education for the end-users.

Schwarz and Morris [8] determined that incorrect and misleading information on websites can have severe consequences for users. The researchers focused on the identification of credibility in web pages to aid users in assessing the validity of the content. This research is critical to our study, as it reinforces the value of flagging malicious websites to notify users. As Bruijn and Janssen stated, the human factor (or socio-technical aspects) of cybersecurity can be the most crucial in preventing a cyberattack [7]. Schwarz and Morris created a visualization of search results to help individuals determine the validity of a website [8]. Providing a tool to allow users to make the right choice when engaging website content [8], works in combination with our research; that users need to comprehend how close they may be to a malicious website. Our model aims to build on this idea by providing users the knowledge of how close they are to a malevolent web page.

## 3 Method

We designed the quantitative study to test a model for identifying risk to websites based on the relationship to known threat indicators. The general problem is that websites can include content from other sites that may load malicious data into a web browser and cause harm to users. The specific problem is known malicious website content can be blocked, but a safe site may load content from a threat which has not yet been labeled with an indicator in the last defense cybersecurity software. The purpose of the research was to describe a method of identifying risks to websites based on the relationship to known threats.

For the study, the independent variable was a threat indicator to a known malicious website. The dependent variable was the relationship between websites inferred by the referring links between all the web traffic captured. For the same window of the captured data, blocked websites were extracted as the threat indicators and mapped as malicious websites. All websites in the captured traffic represented the population for the study and the websites with a relationship to malicious websites represented the sample of the population. Relationships were described as any path from a target website to a known malicious website. The relationships were limited to four "hops" from a website to a malicious site with a threat indicator.

Threat indicators identify malicious websites, but based on the purpose of the study, we sought to test a method of identifying risks to users before they visited a malicious website. Predicting threats to users and averting risk provides a first-move advantage [6] and could reduce the potential risks to an organization. A website that loads content from another website represents a potential link one hop away. If a user clicks on a link to visit a new website, the new website could load data from a third website, which represents a risk two hops away.

A website may contain relationships with multiple malicious sites at multiple hop lengths, so risks could be elevated based on not only the distance of the relationship but also the totality of the relationships. We did not attempt to establish a quantitative sum of the risk in the current study, but the topic is addressed in the future research section. For the study, the following hypothesis and null hypothesis were developed.

H = The relationship based on referrer links and the number of hops to a malicious site can indicate the risk to a website.
$H_0$ = A risk of visiting a website cannot be indicated by the relationship to known threats.

## 3.1 Data Collection and Transformation

Once the research design and method were developed, the extract, transformation, and load (ETL) process began. The data was repurposed from previously collected website traffic [9] and sanitized before ingesting into the databases. The website traffic was gathered from unencrypted network traffic using http-monitor tool [10] and written to a JavaScript Object Notation (JSON) files. We developed a custom Python program to extract the domain, port, URI (Universal Resource Identifier), and referrer data points for ingest into the databases. The referrer represents the URI from the previous website, which leads to the current site. The URI for the original and referrer website was stripped of variables to sanitize any potentially sensitive content. The Python program transformed the content and populated the data into the Neo4j and MySQL databases.

### 3.1.1 Instrumentation

Data collection occurred over 21 days and was limited to the metadata of web traffic occurring on port 80 since metadata from encrypted traffic cannot be extracted. The monitoring tool collected the metadata and web headers of 6,038,580 web traffic connections. We filtered the traffic destined for the websites of the organization to avoid capturing sensitive data. After filtering potentially sensitive data, the number of web traffic connections stored in the MySQL database was 3,735,355. The resulting traffic was stored in MySQL to support queries of domain association generated from the graph database.

Neo4j was selected as the graph database for the research based on the popularity of the software and robust capabilities of existing Python modules. We leveraged the work of the Neomodel module [11] since the object-oriented design allowed for quick code development. All website domains were created as *Website* nodes in the database, and the referring website was also created as a *Website* node. A connection was created between the two nodes using the relationship name of *Refers*. All blocked domains during the time of the web traffic sampling were created as *Indicator* nodes, and a *Threat* relationship was established to the *Website* node. Figure 1 contains the nodes and relationships incorporated within the Neo4j database after the data was loaded.

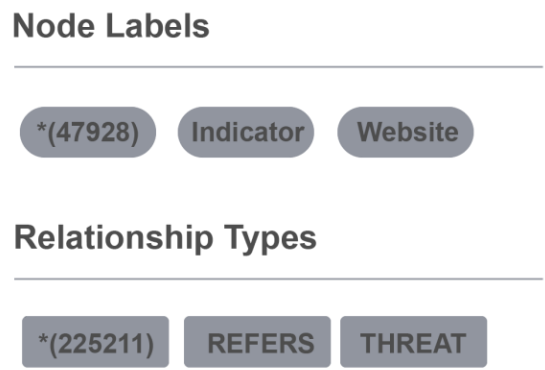

**Figure 1: Neo4j populated nodes and relationships**

With all nodes created and relationships established, we could query the Neo4j database using the cypher query language. The following cypher code provides an example query used during the study to build relationships to malicious websites identified by threat indicator within three hops.

$$MATCH\ path = (:Indicator) - [:THREAT] \rightarrow (:Website) \leftarrow [:REFERS * 1..3] - (:Website)$$
$$RETURN\ path$$

The Neo4j application provides a graphical visualizer capability to render the results of a query. Sample results from the above cypher query are included in Figure 2 to provide a visualization of all possible paths to a malicious website.

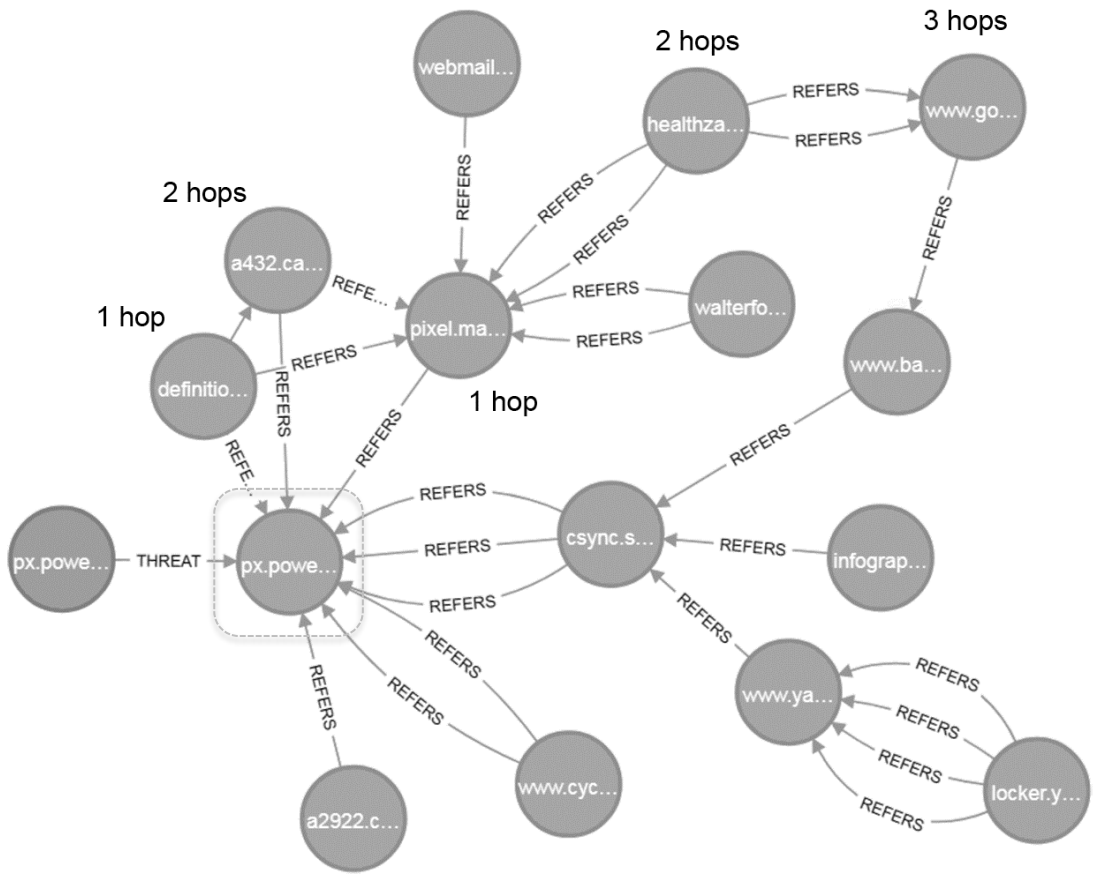

**Figure 2: Referrer hops to a malicious website**

Websites include referring links to other websites, and the hierarchy of associations can be plotted out to any number of "hops." The term hop represents the number of links a user would need to click to arrive at the website under investigation. In Figure 2 the malicious website is labeled as "px.powe" and located on the middle-left with the incoming *Threat* relationship.

As denoted in Figure 2, multiple paths exist to the malicious *Website* node. Two of the paths traverse 3-hop relationships, and seven paths involve a 2-hop relationship via the node denoted such as "csync.s." The shortest number of hops does not denote the expected path a user will traverse, but additional research could test the number of relationships to determine the most likely path. Websites such as "healthza" in the top-right location represent nodes that have multiple paths to the malicious domain via 2-hop and 4-hop relationships.

The cypher query execution identified a malicious *Website* node if it contains an *Indicator* with a *Threat* relationship. The website a user visits is indicated as a node that contains a *Refers* relationship with one to four hops of relationship to the malicious *Website* node. For example, the node on the top-right denoted as "www.go" represents the website the user wants to visit, and the node on the lower left in the figure denoted as "px.powe" with a *Threat* relationship represents the malicious website. Some of the nodes are annotated in the figure with the number of hops away from the desired website. The domain names are not fully annotated to provide some anonymity to the website traffic.

### 3.1.2 Sampling

For the current study, we were interested in associations to websites with known threat indicators. We relied on existing threat indicators from firewall software that blocked traffic to known malicious websites. All metadata for website traffic was captured, but the sampling for the study relied on malicious websites and derived relationships.

While the relationships of the nodes were analyzed within the Neo4j database, we simultaneously loaded the same data into MySQL for alternate analysis methods. The same custom Python script performed the extraction of the data from the source JSON files before inserting it into the Neo4j and MySQL databases simultaneously. Both databases leverage different technical strengths, so Neo4j was used to express relationships, while MySQL was use relied on for the ability to combine and query objects.

## 4 Results

We loaded the captured data into Neo4j to build the relationships, then queried the same data in MySQL to track the paths of users through related websites. The results of the tracked paths of users populated a secondary database table in MySQL to facilitate the final analysis. We queried from the derived table to extract the data needed for the ROC analysis.

Table 1 contains the results of the queried data, which was grouped by the number of hops the website was located from the known malicious site. In the table, False Positive (FP) represents all visitors to related websites who did not visit the final threat. False Negative (FN) represents direct visits to a threat that our model had no way of representing. True Positive (TP) represents all visitors to related websites who also visited the threat. True Negative (TN) represents all visits to websites that were not related to a known threat.

**Table 1: Summary of visits to websites**

| Hop | False Positive | False Negative | True Positive | True Negative |
|---|---|---|---|---|
| 4 | 2406 | 159 | 276 | 6257 |
| 3 | 1884 | 159 | 271 | 6257 |
| 2 | 901 | 159 | 241 | 6257 |
| 1 | 0 | 159 | 298 | 6257 |
| 0 | 474 | 159 | 225 | 6257 |

Note. The data represents the summation of the analysis of traffic visits and the association of websites to known threats. The hop column denotes the distance to a known threat with a hop of zero representing the threat. FP for 0-hop data represents direct visits to non-threat websites without any referring traffic.

Fawcett defined ROC analysis based on calculations from a confusion matrix [2] as depicted in Table 2. Each hop row of data provided input to a single confusion matrix, and the inclusion of all matrices represented the developed model for the current study. Each hop count analysis was evaluated independently, and no analysis was performed on all hop data as a summary of the model, but we provided further analysis of the data in the discussion section.

## Table 2: Confusion matrix [2]

|  | p | n |
|---|---|---|
| Y | True Positives | False Positives |
| N | False Negatives | True Negatives |
| *Column totals:* | P | N |

A confusion matrix contains the four possible outcomes for any instance with the true class represented by the columns and the hypothetical class represented by the rows [2]. For each of the hops in Table 1, we populated a confusion matrix and performed the ROC calculations. The results of the calculations are documented in Table 3. The calculations were performed based on common metrics suggested by Fawcett [2].

## Table 3: Summary of ROC analysis

| Hop | FP Rate | TP Rate | Precision | Sensitivity | Accuracy | F-Measure |
|---|---|---|---|---|---|---|
| 4 | 0.375 | 0.635 | 0.103 | 0.635 | 0.719 | 0.177 |
| 3 | 0.294 | 0.630 | 0.126 | 0.630 | 0.762 | 0.210 |
| 2 | 0.140 | 0.603 | 0.211 | 0.603 | 0.860 | 0.313 |
| 1 | 0.000 | 0.652 | 1.000 | 0.652 | 0.976 | 0.789 |
| 0 | 0.074 | 0.586 | 0.321 | 0.586 | 0.911 | 0.415 |

Note. The data in the table represents the results of the calculations from the confusion matrices. The FP and TP are used to graph the data while the remaining metrics provided the researchers with values to evaluate the certainty of the results.

We did not calculate the negative predictive value for the data since the TN and FN values were static across all hop distances. The values were static since they were based on direct visits to known threats and visits to websites not related to known threats. The value for all hop distances for the negative predictive values was 0.9752. The positive predictive value is denoted in Table 3 as the "Precision" value.

# 5  Discussion

Zweig and Campbell explained that the graph of ROC performance would plot an improved accuracy model as tending toward the upper-left corner [12]. The upper-left corner represents a perfect result with a 100% TP rate and a 0% FP rate. Fawcett further explained that "conservative" classes would tend toward the left side of the graph, while more progressive classes tend toward the right side of the graph [2]. A difference between the two classes is related to the prediction of true positives with a higher willingness to accept increase false positive rates [2]. The graph of the ROC performance for the current study is depicted in Figure 3. The dotted line represents x=y and denotes a model whereby the true positive and false positive rates are equal and, therefore, cannot accurately distinguish between true and false positive values. Predictive results can only provide a useful model by occurring above the dotted line.

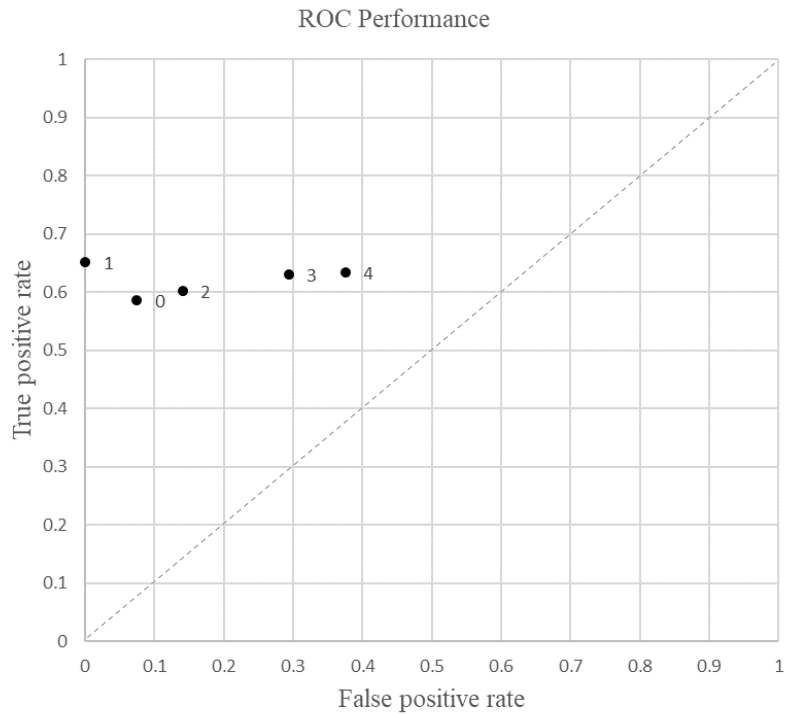

**Figure 3: ROC performance graph based on distance from a threat**

The ROC performance of the 1-hop value does not conform to the pattern of the other hops and is due to the nature of the data collection. The website relationships were generated based on referring links from a known threat and subsequent referring links from derived relationships. Since the first hop from known threats was always visited links, no data could be derived from links to an unknown set. Our model of predicting risk to websites based on association to known threats can still provide value based on the hops of further distances. The issue of inconclusive results for 1-hop values was addressed in the future work section.

The FP rate increased as the number of hops increased, which suggests that the further the relationships are from the original website, the less likely the model can predict threat. The conclusion appears to be valid since the further away a website is from the threat, the more possible paths exist; therefore, the more likely a user will follow an alternate path. The conclusion will not be valid for all websites since some content may provide a higher appeal, which could attract users at an increased rate compared to other content. When considering all websites as a unit, the likelihood of predicting a path to a malicious website appeared to be reduced the further away a relationship was evaluated. More specifically, the false positive rate increased with the distance and may present an unacceptable model for organizations and the users.

The model did not reveal substantial differences in the TP rate as the hop count relationships were increased. The FP rate did increase as the hop count relationship increased; therefore, the data suggests that sacrificing FP rates did not yield a substantial increase in predictive value. Specifically, the FP rate more than doubled from 2-hop to 3-hop relationships. The increased rate of blocked sites could increase user annoyance without providing a significant reduction in threat. Each organization would need to evaluate the user experience compared to threat reduction to determine the proper balance.

The hypothesis for the current study was that the relationship based on referrer links and the number of hops to a malicious site could indicate the risk to a website. The null hypothesis for the current study was that the risk of visiting a website could not be indicated by the relationship to known threats. Since the analysis methods based on the hop distance to a known threat suggested a false positive and true positive rate which exceeded a random guess, the null hypothesis was rejected, and the hypothesis was accepted.

While true positive rates in the study between 0.5859 and 0.6345 may not suggest an outstanding performance, the goal of the study was to develop metrics for enhancing existing capabilities. Cybersecurity products provide indicators to threats, but actions on the indicators are executed singularly without evaluating the relationships. Enhancing the performance of existing indicators with threat levels could provide users with additional information about the potential risk during browsing without necessitating a block of the network activity.

## 6 Future Work

The association and distance to a known threat could be used to develop a method for establishing the risk to non-malicious websites. Researchers could perform a future study to produce a numerical scale that corresponds to the risk to a website based on the number of relationships to known threats. Like Gyongyi et al. definition of an oracle function [4], a website one hop from a malicious website may present a greater risk to a website two hops from a known risk. The calculations would need to assign values to the distance from a risk to determine if, for example, a website with two relationships to a known threat at a distance of two hops is worse than a website with one hop from a known threat.

Once researchers can describe the calculation of risk for websites based on relationships, a scale could be developed to quantify the threat. The scale could limit the results to a known set such as between one and 10 and thresholds be developed for ranges. If the evaluation of the risk model were implemented in a product, an organization could set the threshold for various ranges with actions for each range. For example, threat values below five would not produce any notification to users, but values between five and seven would inject a warning notification that a user was browsing near known threats. Finally, values above seven would trigger a block of traffic since the risk of a user visiting a malicious site was very likely to occur. Each organization could evaluate the scale to determine the acceptable risk based on business needs.

As stated in the discussion section, the method of data collection produced skewed results for 1-hop relationships. To avoid the problem and test the model again, researchers could build the relationships of websites based on a web crawl. The web crawl would process web page content, identify links, and build relationships based on the links versus our method of building relationships based on user web traffic. The links would then be followed to build the relationships of the sites to be studied. This method would require collecting web traffic and sampling only the website visits which corresponded to domains within existing relationships.